\documentclass[10pt,doublecol]{iopart} 
\usepackage[draft]{hyperref} 
\usepackage{iopams}
\usepackage{graphicx}
\usepackage{fancyhdr}
\usepackage{float}
\usepackage{caption}
\usepackage{subcaption}

\begin{document}

\title[Accelerated Thermalisation of $^{39}$K atoms in a Hybrid Trap]{Accelerated Thermalisation of $^{39}$K atoms in a Magnetic Trap with Superimposed Optical Potential}
\author{Dipankar Nath, R. Kollengode Easwaran, G. Rajalakshmi and C.S. Unnikrishnan}

\address{ Tata Institute of Fundamental Research, Mumbai, India}
\eads{\mailto{dip@tifr.res.in}, \mailto{unni@tifr.res.in}}

\pacs{37.10.Gh,67.85.-d,05.20.Dd}

\begin{abstract}
We report the rapid accelerated thermalisation of Potassium ($^{39}$K) atoms loaded in a magnetic trap, in the presence of a single dipole trap beam. More than an order of
magnitude reduction in the thermalisation time, to less than a second, is observed with the focused off-resonant beam occupying only 0.01\% of the volume of the magnetic
trap. A new method for testing for thermalisation of relatively hot clouds is devised. The cold atoms are loaded from a Magneto-Optical Trap (MOT) of $^{39}$K that has gone through a compressed MOT and sub-Doppler cooling stage. The atoms are
prepared in the magnetically stretched $|F=2,m_F=2\rangle$ state prior to loading into the hybrid trap. We also report a direct loading of $^{39}$K atoms, prepared
in the state $|F=1\rangle$, into a single beam dipole trap.
\end{abstract}

\maketitle

\section{Introduction}
Ultra-cold and quantum degenerate bosonic and fermionic gases can serve as a
test bed for various quantum mechanical as well as quantum statistical
phenomena. They have also been important in developing various applications
and in studies in fundamental physics. A first step towards producing a
degenerate gas of atoms is to load the pre-cooled atoms from a Magneto Optical
Trap (MOT)\cite{ELR:1987} into a conservative trap like a pure Magnetic
Trap\cite{DEP:1983,CJM:1997,TE:1998} or an Optical Dipole Trap\cite{MDB:2001}.
In order to attain lower temperatures, evaporative cooling is performed on the
cloud of atoms in the conservative trap. Dipole trap has the advantage that the
atoms need not be spin polarized and can be prepared in any one of the
hyperfine ground state. However dipole traps inherently have a smaller volume
in comparison with a magnetic trap. Also, effective 3-D confinement is
possible only with a Crossed Dipole Trap (CDT).

The combination of a quadrupole magnetic trap and an optical dipole
trap\cite{YJL:2009}, however provides us with the merits of both kinds of
confinement. The center of the dipole trap (i.e. the focus of the beam) is
overlapped with the center of the quadrupole trap. To prevent atom loss and
heating due to Majorana spin flips the dipole trap is kept at a distance equal
to its beam waist from the center of the quadrupole field. The hybrid trap
formed by this combination eliminates the need for a CDT as 3-D confinement is
provided by the combination of the off-resonance optical field and the
magnetic field.

Of atoms in the alkali species, Potassium is a good option for studies
involving both bosons and fermions, in spite of the difficulty in reaching low
temperatures in a vapor loaded MOT. Due to the closely spaced upper state
hyperfine levels of $^{39}$K, the temperature attainable in a $^{39}$K MOT is
larger than 1mK, which is well above the Doppler temperature and a short
molasses cooling phase is introduced to further reduce the temperature lower than
100 $\mu$K \cite{CFO:1998,VGO:2011,MLA:2011}. Loading into a magnetic trap
results in unthermalised sample with relatively high temperatures, and
thermalisation times are typically large, about 15-20 seconds\cite{GMT:1998}.
Direct evaporative cooling is restricted by the fact that the scattering
length is negative and thermalisation is inefficient. It is possible to load
the atoms into a hybrid trap and then switch off the magnetic field, leaving
the atoms only in the Optical Dipole trap. Feshbach resonances can then be
used to tune the scattering length between the atoms\cite{CDE:2007} for
efficient direct evaporative cooling in the optical dipole trap, leading to a
BEC\cite{MLA:2012}. $^{39}$K BEC was first obtained by sympathetic cooling with $^{87}$Rb and
using Feshbach resonance to tune the interaction\cite{GR:2007}.
Recently, direct loading of $^{39}$K atoms into a crossed dipole trap has also
been reported\cite{BSM:2012}. Since Feshbach resonances are forbidden for
atoms in state $F=2$\cite{ABA:2002}, it is essential to prepare the atoms in the state $F=1$ to tune the scattering length, for evaporative
cooling to degeneracy.

In this paper we report two results of importance in this context for
obtaining quantum degenerate samples, a) optical potential assisted rapid
thermalisation in a magnetic trap, resulting in more than an order of
magnitude reduction in the thermalisation time, and b) direct loading of cold
atoms from the MOT, prepared in the state F=1, into a single beam optical
dipole trap. Efficient thermalisation happens in the hybrid trap in spite of
the fact that the volume of the optical trap relative to the magnetic trap is
only about $10^{-4}$.

For the thermalisation experiments, we follow a method similar to
\cite{MLA:2012} while loading $^{39}$K. The cold atoms obtained from the MOT
after sub Doppler cooling are first prepared in the state $|F,m_{F}%
\rangle=|2,2\rangle$. The atoms are then transferred to a quadrupole trap
after all the MOT beams are switched off. An optical dipole beam (1080nm,
20W-50W) with its center overlapped with the center of the quadrupole field is
also switched on. In this way we can transfer almost 80\% of the MOT atoms
into the hybrid trap.

In order to directly load $^{39}$K atoms from the MOT to a single beam dipole
trap in $F=1$ state after sub-Doppler cooling stage, the dipole trap beam is
switched on after the MOT beams and magnetic field are switched off, after a
short phase of optical pumping into the $F=1$ state with only the cooling beam
on and the repumper off. About $10^{5}$ atoms are captured in the single beam
dipole trap.

\section{Experiment}
\subsection{Magneto Optical Trap}
\label{subsec:MOT} The vacuum assembly of the set up is described in
\cite{VGO:2011}. Water cooled coils are used to create the inhomogeneous
magnetic field required for the MOT and later the quadrupole magnetic trap.
The atoms in the 3D MOT are loaded from a 2D$^{+}$ MOT\cite{KDI:1998}. Due to
the small level spacing in the hyperfine structure of $^{39}$K, both the
cooling $\left( F=2\rightarrow F^{\prime}=3\right) $ and repump $\left(
F=1\rightarrow F^{\prime}=2\right) $ beams are detuned below the entire
excited state hyperfine levels. As shown in \cite{ABA:1997}, the cooling
efficiency is different for different combination of intensity and detuning.
For loading the MOT, both the cooling and repump beams are red detuned with
respect to the entire hyperfine structure of the excited states and the
intensity is kept high. This helps in capturing a large number of atoms in the
MOT but the cooling is ineffective. Temperature attained is $\geq1m$K. If the
cooling and repump beams are slightly red detuned only with respect to the
respective hyperfine resonances and the intensity is kept small, the cooling
is efficient but less number of atoms are captured. Thus a two stage cooling strategy is adopted in order to trap a large
number of atoms at a lower temperature. First, the MOT is prepared with beams of high
intensity and detuning.  To cool the atoms further a sub-Doppler stage is
introduced, where both the intensity and the detuning is decreased
\cite{VGO:2011,MLA:2011}. Temperatures of the order of 25-30$\mu$K have been reported.

 Prior to transferring the atoms into the magnetic
trap, the atoms are first compressed using a combination of increased magnetic
field gradient as well as change of the detuning and intensity of the laser
beams. This stage is called the Compressed MOT(C-MOT) stage (See Table
\ref{tab:c-mot}).

\begin{table}[h]
\caption{Values of various parameters during C-MOT stage. $\delta_{C}%
,\delta_{R}$ are the cooling and repump detuning,$I_{C}$, $I_{R} $ and $I_{S}$
are the cooling, repump and the saturation intensity respectively}%
\label{tab:c-mot}%
\vskip 0.5cm \centering
\begin{tabular}
[c]{cccccc}\hline\hline
& $\frac{\delta_{C}}{\Gamma}$ & $\frac{\delta_{R}}{\Gamma}$ & $\frac{I_{C}%
}{I_{S}}$ & $\frac{I_{R}}{I_{S}}$ & $B^{\prime}$\\
&  &  &  &  & (Gauss/cm)\\\hline\hline
Initial Value & 4 & 2.3 & 2 & 5.5 & 8\\
Final Value & 7 & 2.3 & 2.7 & 5.5 & 20\\\hline
\end{tabular}
\end{table}After the C-MOT stage, a molasses cooling stage is implemented in
which we change the intensity and detuning of both the cooling and repump
beams. The values of the intensity and detuning during the sub Doppler cooling
stage are shown in Table \ref{tab:subdoppler}. \begin{table}[h]
\caption{Values of various parameters during sub Doppler cooling stage}%
\label{tab:subdoppler}%
\vskip 0.5cm \centering
\begin{tabular}
[c]{cccccc}\hline\hline
& $\frac{\delta_{C}}{\Gamma}$ & $\frac{\delta_{R}}{\Gamma}$ & $\frac{I_{C}%
}{I_{S}}$ & $\frac{I_{R}}{I_{S}}$ & $B^{\prime}$\\
&  &  &  &  & (Gauss/cm)\\\hline\hline
Initial Value & 7 & 2.3 & 2.7 & 5.5 & 20\\
Final Value & 1.5 & 1.9 & 0.4 & 0.2 & 0\\\hline
\end{tabular}
\end{table}Temperature of the order of 70$\mu$K is achieved after the two
stages described above.
\subsection{Loading into a magnetic trap}
In order to load the atoms into the quadrupole magnetic
trap, the atoms have to be prepared in the magnetically stretched
$|F=2,m_{F}=2\rangle$ state, by shining a circularly
polarized light beam slightly blue detuned to the transition $F=2
\rightarrow F'=2$ on the atoms in the MOT. A magnetic field of around 5 Gauss is also switched
on. To prevent optical pumping to the hyperfine level $F=1$ we also mix a beam
resonant with the repump transition along with the spin polarizing beam. In order to avoid heating during the spin polarization process, we keep the state preparation beams on for a short duration of about 100~$\mu$s only. This is very essential for the loading of atoms into the magnetic trap as long duration pulses reduce the efficiency with which the atoms are captured in the trap. The
quadrupole magnetic field is switched on immediately after preparation of the
atomic cloud in the desired state. It takes about 10ms to reach the final
magnetic field gradient of about 95 Gauss/cm in the plane perpendicular to the coil axis (190 Gauss/cm along the coil axis). Loading efficiency of
about 80\% has been observed. The atoms in the magnetic trap are imaged via absorption imaging in
the 2f-2f configuration, using an absorption probe resonant to the cooling transition ($F=2\rightarrow F^{\prime}=3$).

\subsection{Hybrid Trap and Dipole Trap}
The hybrid trap is created by aligning the dipole trap center (the focus of the
dipole trap beam) with the center of the quadrupole trap. The dipole trap is
derived from a 1080~nm ($\Delta\lambda\sim$~1nm) unpolarized fiber laser (Manlight,\ France).

The dipole beam has a beam waist of 25~$\mu$m at the focus and a
Rayleigh length of about 1800~$\mu$m, which exceeds the magnetic trap size. At
a maximum power of 50 W, the trap depth of the dipole trap is around $1.9$~mK.
The dipole trap beam is switched on together with the magnetic trap. The
effective potential of the Hybrid trap is given by\cite{YJL:2009}:
\begin{eqnarray}
U\left(  \vec{r}\right)   & =\mu B^{\prime}\left(  x^{2}/4+y^{2}+z^{2}/4\right)  ^{\frac{1}{2}}\nonumber\\
& -U_{0}exp\left(  -2\left(  y^{2}+z^{2}\right)  /w_{0}^{2}\right) \nonumber\\
& +mgy\label{eqn:hybridtrappotential}
\end{eqnarray}
where $\mu$, $B^{\prime}$, $U_{0}$ and $g$ are the magnetic dipole moment of
the atom, magnetic field gradient, dipole trap depth and the acceleration due
to gravity respectively. The dipole trap is along the x-axis. Here, we are considering the case where the focus of
the dipole trap beam coincides with the center of the magnetic trap. Typical dipole trap depth and magnetic field gradient used in our experiment are 
$U_{0}=1.9$~mK and $B^{\prime}=$190~Gauss/cm.

\section{Results}

\subsection{Thermal Relaxation in Hybrid Trap}

\begin{figure}[h]
\centering
\includegraphics[width=.45\textwidth]{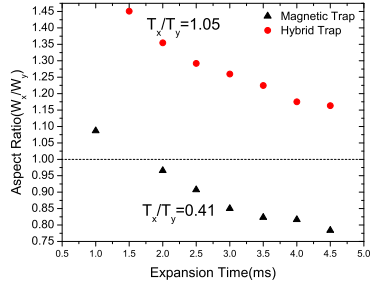}\caption{The aspect ratio $W_x/W_y$, of the atomic cloud for various expansion times after releasing from the Quadrupole Magnetic Trap and Hybrid Trap with a holding time of 2~s. $W_x$ and $W_y$ are the cloud size along the $x$ and $y$ axis respectively}
\label{fig:aspectratioasfunctionofexpansion}%
\end{figure}
The transfer of atoms from the MOT to the magnetic trap leads to an unequal distribution of energy (temperature) and hence expansion velocities along different directions. This anisotropy and its time evolution can be measured from the rates of expansion of the atomic cloud along different directions, after releasing the atoms from the trap. During free expansion, the size of the cloud increases with time $t$ in the two directions, $ i=x,y $, as $ {W_i}^2 ={W_{i0}}^2+{v_i}^2t^2 $ where $v_i$ is the expansion velocity and $ W_{i0} $ is the initial size of the atomic cloud in the trap. The effective temperature along any one of the axis, $ T_i\propto {v_i}^2 $.  In one set of experiments the temperature was measured along the shallow (x) as well as the tighter (y) direction after 2s of holding in both the magnetic quadrupole trap and the hybrid trap. In the magnetic trap, we measure a large initial temperature anisotropy $ T_x/T_y \simeq 0.4 $, with a very slow thermal relaxation rate, estimated to take 
about 20 s (see below). In the hybrid trap, in stark contrast, the cloud is already well-thermalised within the holding time of 2 s, with $ T_x/T_y \simeq 1 $. The ratio of the sizes in the two directions (the aspect ratio) as a function of duration of expansion, for both cases, is plotted in Figure \ref{fig:aspectratioasfunctionofexpansion}. The clear difference in the aspect ratios in the two cases allows us to use this as a test for thermalisation. 

A standard way to observe thermal relaxation would be to allow for a long expansion duration after which the aspect ratio for a thermally relaxed cloud would approach unity as the expanded cloud size will become independent of the initial cloud size. This is feasible for samples with temperatures well below 100~$\mu$K and initial size occupying a small fraction of the field of view. However for samples with temperatures well above the Doppler temperature, the thermal expansion is too fast for this technique to be reliable. Hence we have devised a new technique to test for thermalisation with relatively small duration of expansion.

To measure the thermalisation time, we let the cloud expand
after varying amounts of holding time in the trap. After a long expansion
time, a thermally relaxed cloud attains an aspect ratio of 1 asymptotically.
For a non-thermal cloud, \emph{the aspect ratio crosses unity during the
expansion} because the expansion rate is different along different directions (Figure \ref{fig:aspectratioasfunctionofexpansion}). Typically the temperature and the expansion rate in the direction of tighter trapping is larger. This implies that for thermally relaxed cloud expanding for a fixed amount of
time, the aspect ratio should be independent of the holding time in the trap
once the holding time exceeds the thermalisation time. If the holding time is
lower than the relaxation time, the expansion rate in the two directions will
remain different. 

\begin{figure}[h]
\centering
\includegraphics[width=.45\textwidth]{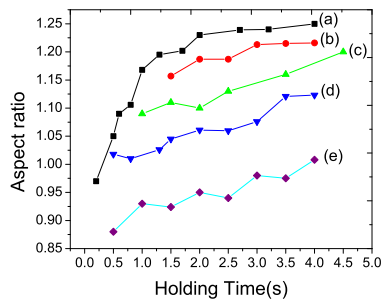}\caption{(a)
shows thermal relaxation in a hybrid trap, (b),(c) and (d) shows the thermal
relaxation in magnetic trap after the dipole trap has been kept on for 1s,
500ms and 200 ms respectively, (e) shows the thermal relaxation in the
magnetic quadrupole trap only.}%
\label{fig:thermal_relaxation}%
\end{figure}

We measure the aspect ratio of the cloud after allowing for 4ms of expansion for various holding times. It is observed that beyond a certain holding time $\tau_{c}$ in the hybrid trap, the aspect ratio becomes nearly independent of the holding time (see Figure \ref{fig:thermal_relaxation}). We call the time required to approach to 90\% of the final aspect ratio, with anisotropy in temperature less than 5\%, as the thermal relaxation time $\tau_{c}$ in the hybrid trap.

From Figure \ref{fig:thermal_relaxation} we can see that the thermal
relaxation in case of the hybrid trap is an order of magnitude faster than that
of the quadrupole trap. In order to confirm that the accelerated thermal
relaxation is solely due to presence of the dipole trap beam, we keep the beam
on for 1~s, 500~ms and 200~ms and allow the cloud to relax in the magnetic
quadrupole trap only. We see that cloud continues to relax at a slower pace
(as in the case of the magnetic trap) once the dipole trap is switched off. We
measure the thermal relaxation time in the hybrid trap to be about $\sim
0.7\pm.09$s.

The relaxation time in the pure magnetic trap is too long to be measured
accurately. We can see only the linear part of the slow thermal
relaxation in the case of magnetic trap in Figure \ref{fig:thermal_relaxation}. The thermal relaxation has been obtained approximately by fitting it to a
linear function and is found to be more than 20~s, consistent with previously
reported long relaxation times in a quadrupole magnetic trap of more than 10 s
\cite{GMT:1998}. Theoretical calculation of the scattering length \cite{MLAThesis:2012} suggests that the Ramsaeur minimum is around 400 $ \mu $K. However experimental studies of the scattering length variation with temperature are very few and the measured numbers differ from the calculated values. This calls for a more careful experimental determination of the Ramsaeur minimum.

The physical mechanism for relaxation towards thermal equilibrium is
collisions. Atoms need to undergo sufficient number of collisions in the
hybrid trap to redistribute the energy equally among the motional degrees of
freedom in different directions and become thermally relaxed. The relaxation
time for thermal equilibrium is of the order of $3\gamma_{c}^{-1}$
\cite{KBD:1995,JDA:1998} where $\gamma_{c}$\footnote{$\gamma_{c}=\bar{n}%
\bar{v}\sigma$.where $\bar{n}$ is the average density, $\bar{v}=\sqrt
{\frac{8k_{B}T}{\pi M}}$ is the average velocity and $\sigma$ is the collision
cross-section} is the collision rate per atom. However, the dipole trap beam
occupies only about 0.01\% of the total hybrid trap volume and yet,
essentially all the atoms in the hybrid trap become thermally relaxed.
Therefore, thermalisation should happen during the multiple crossing passages
of the atoms in the magnetic trap through the single dipole beam that spans
the entire length of the weakly trapping direction of the hybrid trap. The
fact that atoms spend only small fraction of time (less than 1/10 of the 700 ms we
measure) within the dipole beam regions accentuates the dramatic reduction in
the relaxation time in the presence of the dipole beam.

The typical collision rates can be estimated from the approximately measured
atom number densities in different regions of the hybrid trap. We estimate the average
atomic density ($\bar{n}$) of the atoms in the dipole trap region, $\bar
{n}_{dipole}$ $\approx$ 90 $\times$ $\bar{n}_{magnetic}$ (density in magnetic
trap). This induces a collision rate ($\gamma_{c}$) of 48/s at the dipole trap
region compared to 0.5/s in the magnetic trap. Thus almost all the collisions
required for thermal relaxation occurs in the dipole trap region. We estimate
that within the measured thermal relaxation time $\tau_{c}\approx
700$~ms the atoms spend less than 40\thinspace ms on the average
inside the dipole trap. This time is evaluated for our magnetic trap
oscillation frequency of 200\thinspace Hz and for velocity of the atoms
$\approx$ 0.56~m/s (for 500$\mu$K atomic cloud). Thus each atom undergoes at
most 2 collisions during the time it spends in the dipole trap. However, the maximum velocity
of the atoms is almost a factor of 3 larger as the atom traverses the 1.9 mK
deep optical potential well, and the effective collision rate is expected to
be proportionately higher. This is marginally consistent with the number of
collisions required for complete thermalisation, which is about 5 collisions\cite{DWS:1989}.
The small discrepancy might be due to an under estimation of the atomic
density in the optical dipole trap region. In any case, the vastly accelerated
thermalisation becomes very useful for performing RF evaporation in the hybrid
trap to get ultra-cold atomic samples.

\begin{figure}[h]
\centering
\includegraphics[width=0.45\textwidth]{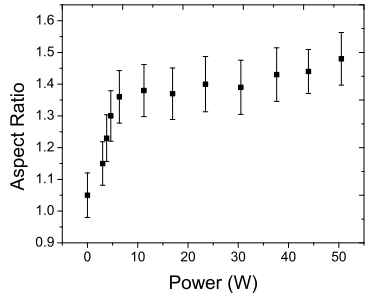}\caption{Aspect
ratio as function of power}%
\label{fig:aspectratiovspower}%
\end{figure}
\begin{figure}[h]
\centering
\begin{subfigure}[b]{0.2\textwidth}
\centering
\includegraphics[width=\textwidth]{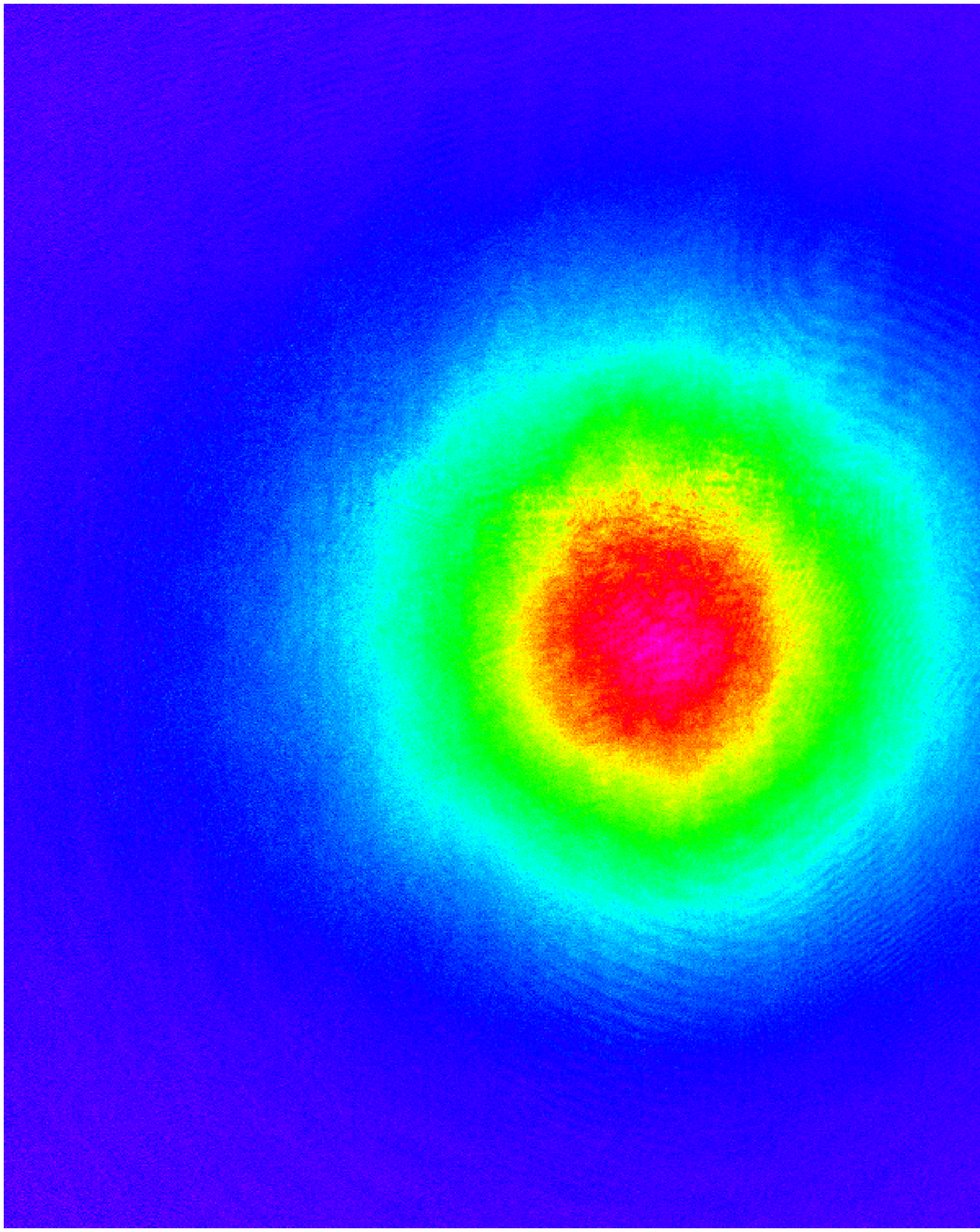}
\subcaption{Quadrupole Trap\\($T_{avg}=375\mu$K)}
\label{fig:quadrupole-trap-2s}
\end{subfigure}
\begin{subfigure}[b]{0.2\textwidth}
\centering
\includegraphics[width=\textwidth]{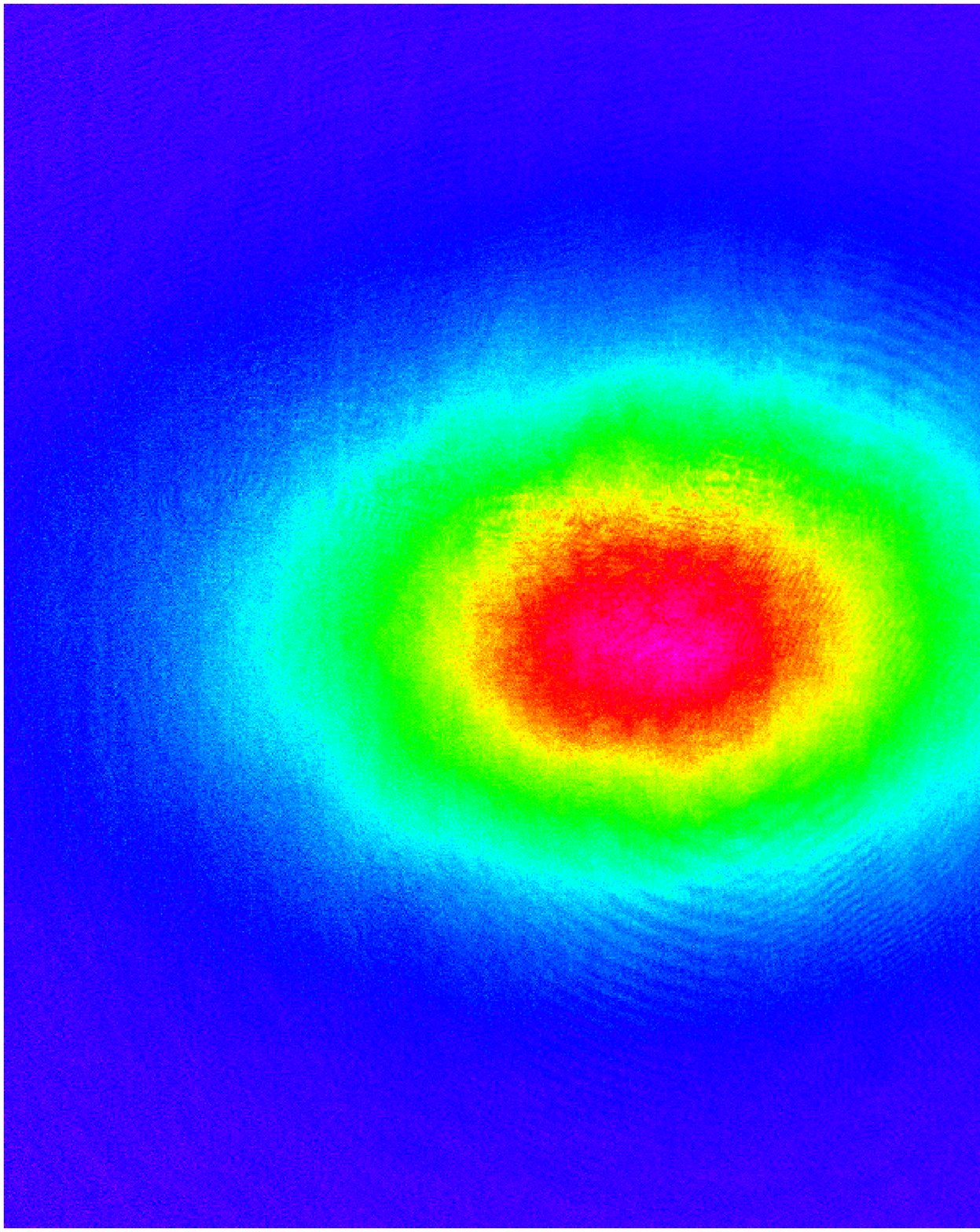}
\subcaption{Hybrid Trap\\($T_{avg}=500\mu$K)}
\label{fig:hybrid-trap-2s}
\end{subfigure}
\caption{Quadrupole Trap and Hybrid Trap.}%
\label{fig:quadrupole-hybrid-trap}%
\end{figure}

For a fixed holding time, the thermal relaxation will depend on
the power in the dipole trap. We hold the atoms in the hybrid trap for about
2\thinspace s and allow the cloud to expand for 1.5 ms before we take the image. The aspect ratio of the cloud is
measured for different values of the dipole trap power. From Figure
\ref{fig:aspectratiovspower} we can see that the aspect ratio starts becoming
steady at higher power. This is because thermalisation takes place within the
holding time and aspect ratio of the cloud in the trap is maintained. The
absorption images of the magnetic trap and the hybrid trap, after a holding
time of 2\thinspace s, are seen in Figure \ref{fig:quadrupole-trap-2s} and
\ref{fig:hybrid-trap-2s} respectively. It must
be noted that the ellipticity of the cloud is due to the anisotropy of the
magnetic quadrupole trap whose gradient is stronger along the axis of the
anti-Helmholtz coil.
The optical potential that occupies a very small relative volume does not
significantly affect the overall shape and ellipticity of the hybrid trap. The near spherical shape of the cloud in Figure \ref{fig:quadrupole-trap-2s} is due to the washing out of the original ellipticity by the anisotropic expansion. The initial size of the magnetic trap is smaller than that of the hybrid trap because of the difference in the average temperatures of the trapped atoms.

\subsection{Loading of atoms into the single beam dipole trap}
\begin{figure}[h]
 \centering
 \begin{subfigure}[b]{.45\textwidth}
 \includegraphics[width=\textwidth]{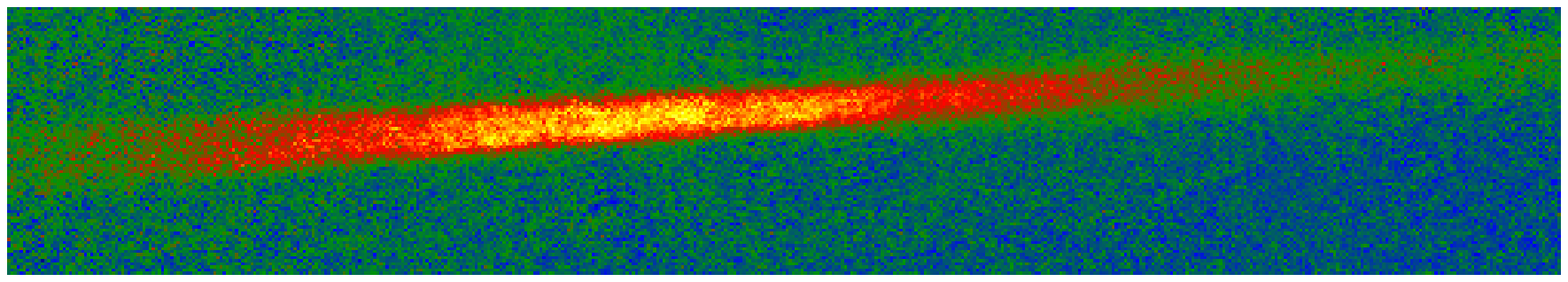}
 \caption{Absorption image of a single beam dipole trap.}
 \label{fig:dipole-trap-absorption}
 \end{subfigure}
 
 \begin{subfigure}[b]{.45\textwidth}
  \centering
  \includegraphics[width=\textwidth]{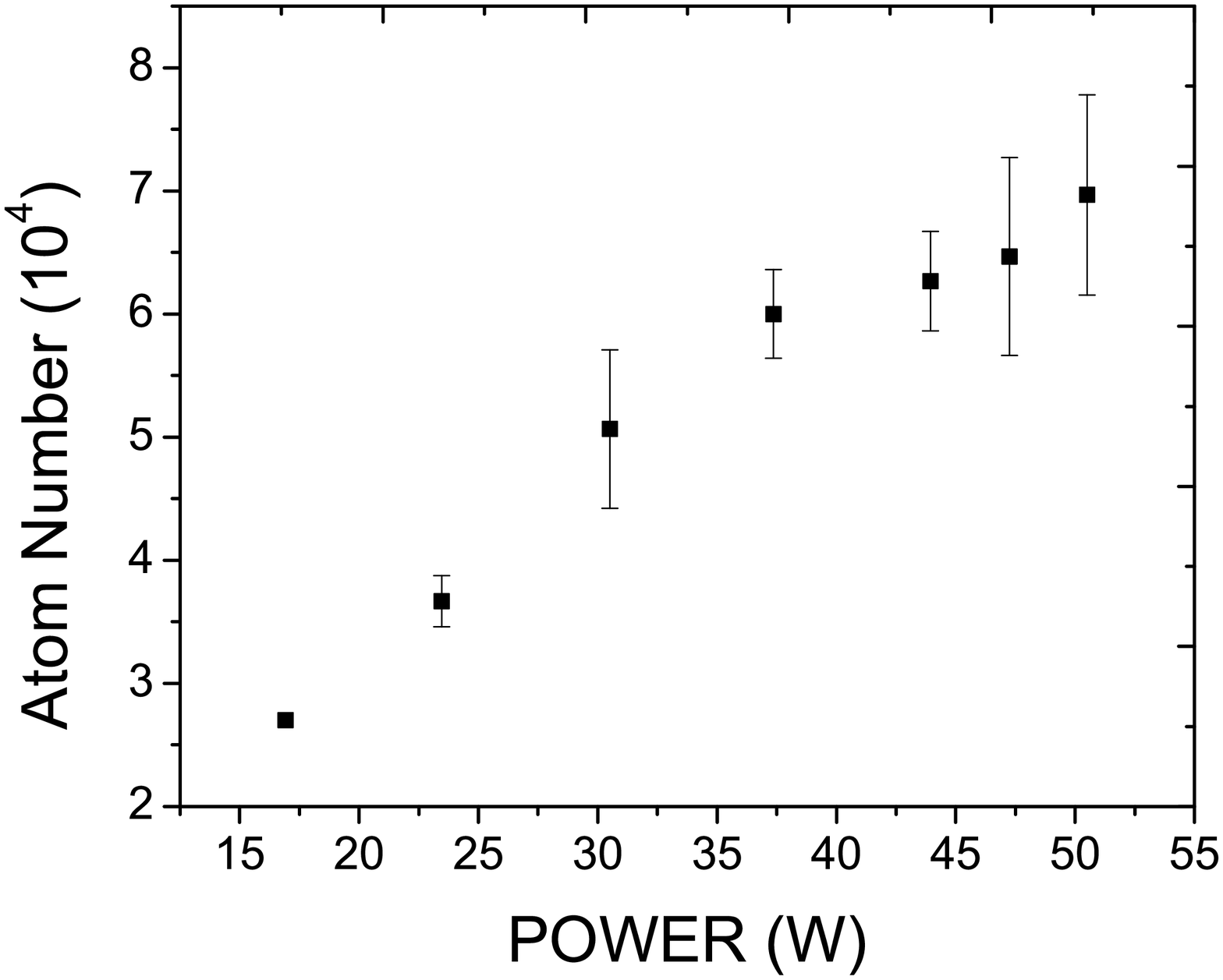}
  \caption{Atom number in single beam dipole trap as a function of power}
  \label{fig:atom-number-vs-dipole-trap-power}
  \end{subfigure}
  \caption{}
 \end{figure}
To load the atoms into the single beam dipole trap, directly from the MOT the atoms are prepared in the $F=1$ state by letting the cooling beam remain on for 1ms after the sub Doppler cooling
stage while the repump beam is switched off. The dipole trap beam is then switched on and atoms are held in the dipole trap for about 50ms after which we take
an absorption image of the trapped cloud. The dipole beam has a beam waist of $25\mu$m at the focus and a
Rayleigh length of about $1800\mu$m.
Figure \ref{fig:dipole-trap-absorption} shows the absorption images of a single beam dipole trap of $^{39}$K. The number of atoms loaded into the single beam
dipole trap is measured as a function of the dipole trap beam power. From Figure \ref{fig:atom-number-vs-dipole-trap-power} we see that the number of trapped atoms 
increases with increasing power. This is different from the situation observed in direct loading into a CDT\cite{BSM:2012}. Temperature of the atoms in the dipole trap is measured to be $250\pm9\mu$K.
\section{Summary}
Accelerated thermalisation by about a factor of 30 has been observed in a cloud of trapped $^{39}$K atoms. This increase is assisted by the presence of a steep 1D optical potential spanning almost the entire diameter of the magnetic trap,
whose volume is smaller by a factor of $10^{-4}$ compared to the
volume of the magnetic trap. Faster RF evaporative cooling of Potassium
as well as other species can thus be performed in such a trap. Direct loading of the $^{39}$K from a MOT into a single beam dipole trap has also been observed.
Prior to loading, the atoms are prepared in the $F=1$ hyperfine state which is conducive for Feshbach resonances. A CDT will be implemented and Feshbach resonances
will be used on the trapped atoms to tune their interaction for efficient evaporation to quantum degeneracy in the dipole trap. 

\section{Acknowledgement}
We would like to thank Mr S.K. Guram for technical support. We would also like to thank Sanjukta Roy for valuable discussions.

\bibliographystyle{iopart-num}
\bibliography{ref}



\end{document}